Phantom chain simulations for the effect of stoichiometry on the fracture of star-polymer networks

*Yuichi Masubuchi

Department of Materials Physics, Nagoya University,

Nagoya 4648603, JAPAN

*correspondence should be addressed, e-mail: mas@mp.pse.nagoya-u.ac.jp

**ABSTRACT**

By phantom chain simulations, it has been recently discovered that the fracture characteristics of star polymer networks with different node functionality and conversion ratios can be described by the cycle rank of the networks [Masubuchi et al., Macromolecules, doi: 10.1021/acs.macromol.3c01291]. However, due to the employed simplifications and idealizations of the examined model networks, the results cannot be cast into realistic systems straightforwardly. For instance, the equimolar reaction was assumed in a limited volume for the binary mixture of star prepolymers. For this issue, the present study investigated the effects of stoichiometry by phantom chain simulations. Examined polymer networks were created from binary mixtures of star prepolymers with various mixing ratios by the end-linking reaction via Brownian dynamics simulations. The networks were stretched with energy minimization until the break. From the mechanical response, strain and stress at break and work for fracture were obtained. These fracture characteristics slightly decrease with increasing the contrast of volume fractions of the binary prepolymer blends when the node functionality is small, and the conversion ratio is large. For the other cases, the stoichiometry does not impact the fracture behavior. The number ratio of broken bonds and the cycle rank exhibit similar stoichiometry dependence. Consequently, the stoichiometry of prepolymer blends does not disturb the previously reported relationships between the fracture characteristics and the cycle rank.

**INTRODUCTION**

Despite lots of attempts, the effect of node functionality $f$ on the network fracture has yet to be fully elucidated [1–3]. The widely adopted approach [4–9] to improving network toughness employs multi-functional network nodes with $f > 4$. However, a few studies [10–12] contradict such a direction. For example, a study on hydrogels made from binary mixtures of star polymers demonstrated that the gels consisting of 3-arm stars (i.e., $f = 3$) exhibit superior fracture properties than their 4-arm ($f = 4$) analogs[10]. A simulation study [12] reported consistent results.

A recent simulation study[13] revealed that the abovementioned contradiction is probably due to the conversion ratio $\varphi_c$; for networks with large functionality (large-$f$ networks with $5 \leq f \leq 8$), the fracture properties are almost insensitive to $\varphi_c$ in $\varphi_c \geq 0.6$, whereas small-$f$ cases ($f = 3$ and 4) they strongly depend on $\varphi_c$. Consequently, at low $\varphi_c$, large-$f$ networks are superior to small-$f$ analogs, but small-$f$ networks become superior to large-$f$ ones at large $\varphi_c$. The superiority of small-$f$ networks at high $\varphi_c$ is explained by the so-called strand extenders that are prepolymers with only 2-arms reacted.

Further interestingly, the simulation results suggest that the cycle rank of the networks $\xi$ dominates the fracture properties. Namely, strain at the break $\varepsilon_b$ for various networks with different $f$ and $\varphi_c$ lie on a master curve if $\varepsilon_b$ is plotted against $\xi$. Stress at the break $\sigma_b$ and work for fracture $W_b$ normalized by the fraction of broken strands $\varphi_{bb}$ are also on master curves concerning $\xi$.

Although the simulation results mentioned above may give some insights into the structure-property relationship of network polymers, mapping to realistic systems is not straightforward due to the employed simplified model. Consistent with the earlier experiments [14–19], the simulations were considered for binary mixtures of star polymers. This clever setting is known to prohibit the formation of primary and odd loops. In addition, the simulation artificially prohibits secondary loop formation, which may affect the mechanical properties [20–22]. It is also noted that the simulations were conducted for exact stoichiometric conditions within small simulation boxes filled with limited numbers of prepolymers. Such mixing means high homogeneity down to each molecule. Since these settings match the mean-field assumption [23–25], the cycle rank of the obtained networks was in excellent agreement with the theory. However, the simulation may realize extremely idealized network structures different from realistic materials.

This study investigated the effects of a few simulation settings on the relation between fracture properties and the cycle rank. Namely, binary star-polymer mixtures with various mixing ratios were employed. Gelation simulations were conducted for such mixtures of phantom chains without prohibiting secondary loop formations. Afterward, the resultant networks were stretched until the break, and the fracture properties were obtained from the stress-strain curves. The results revealed that the fracture properties and the cycle rank are practically insensitive to the examined settings, implying that correlations between end-linking reactions on different arms are insignificant, even in conditions deviating from the stoichiometric ones. Details are shown below.

**MODEL AND SIMULATIONS**

Because the employed model and the simulation scheme are the same as in the previous reports [12,13], readers familiar with those studies may skip this section. Phantom star chains are dispersed in a simulation box with periodic boundary conditions. Each star chain has $f$-arms diverging from the central bead, and each arm has several beads connected by non-linear springs. Using this non-linear spring avoids bond elongation beyond the critical length of bond scission considered in the stretching step. Before gelation, star polymer sols were sufficiently equilibrated by the Brownian dynamics scheme, where the position of each bead obeys a Langevin equation of motion. Afterward, an end-linking reaction was turned on[26,27]. Even though the star polymers are identical, they were colored in two different chemistries with a specific mixing ratio $\varphi_m$. The reaction took place only between end segments with different chemistries. During the gelation process, snapshots of the system were stored when $\varphi_c$ reached specific values. Here, $\varphi_c$ is defined as the ratio of reacted end segments in the system. According to these processes, network structures with various $(f, \varphi_m, \varphi_c)$ values were provided. The networks were stretched by alternatively applying stepwise strain and energy minimization while the Brownian motion was turned off. The bond was severed during the stretch when the length exceeded a specific value. The stretch process was continued until the bond scission violated the percolation.

Below, simulation details are described. The beads number per arm $N_a$ was fixed at 5. The arm number $f$ was varied from 3 to 8. The spring constant is written as $f_{ik} = (1 - \mathbf{b}_{ik}^2/b_{\max}^2)^{-1}$, where $\mathbf{b}_{ik}$ is the bond vector between consecutive beads $i$ and $k$, and $b_{\max}$ is the maximum bond length chosen at 2. The bead number density was chosen at 8, corresponding to the concentration $c/c^* \sim 4$ concerning the overlapping concentration $c^*$. No strong density fluctuation was observed even after gelation with this concentration. In most cases, the total number of star prepolymers $M$ was fixed at 1600. The number fraction of the major component was varied in the range $0.5 \leq \varphi_m \leq 0.6$. This $\varphi_m$ range corresponds to the ratio between major/minor components of $1 \leq \varphi_m/(1-\varphi_m) \leq 1.5$. The Brownian dynamics was numerically integrated with a second order scheme[28] with the step size $\Delta t = 0.01$. For the end-linking reaction, the critical distance and the reaction rate were chosen at $r_c = 0.5$ and $p = 0.1$. The snapshots during the gelation were stored at $\varphi_c = 0.6, 0.7,$ and $0.8$. To the obtained networks, energy minimization was imposed with the Broyden-Fletcher-Goldfarb-Sanno method.[29] The energy conversion and the bead displacement parameters were $\Delta u = 10^{-4}$ and $\Delta r = 10^{-2}$. The magnitude of stepwise uniaxial strain was 0.1. The critical bond length for scission was $b_c/b_0 = \sqrt{1.5}$, where $b_0$ is the average bond length under equilibrium (in the Brownian dynamics simulations for sols). For each condition of the given set of $(f, \varphi_m, \varphi_c)$, eight independent simulation runs were performed for statistics.

**RESULTS AND DISCUSSION**

Figure 1 exhibits stress-strain curves obtained for $(f, \varphi_c) = (4, 0.8)$ with various $\varphi_m$. As reported previously, the stress-strain curves are similar to experiments. Namely, stress increases with increasing strain due to network elasticity, and the non-linearity gradually appears as the applied strain becomes large. Further strain increase induces bond scission, and stress rapidly decreases, reflecting a cascade of bond breakage. Note that if the Brownian motion is considered, the stress decay induced by the bond breakage depends on the dynamics of the network and the strain rate. [12,26] The employed energy minimization scheme avoids such technical difficulties, although the effect of dynamics cannot be discussed.

Nevertheless, comparison among the curves tells that as $\varphi_m/(1 - \varphi_m)$ increases and the mixing condition deviates from the equimolar condition, the network becomes mechanically weak when $f$ is small and $\varphi_m$ is relatively large. In Figure 2, strain and stress at break ($\varepsilon_b$ and $\sigma_b$) and work for fracture $W_b$ extracted from the mechanical response are plotted against $\varphi_m/(1 - \varphi_m)$ for a few $\varphi_c$ cases with $f = 3$ and 8. For the results of $f = 3$ shown in the left column, as $\varphi_m$ increases, $\varepsilon_b$, $\sigma_b$, and $W_b$ decreases at $\varphi_c = 0.7$ and 0.8 (triangle and cross). In particular, $W_b$ for $\varphi_c = 0.8$ (red cross in panel c-1) is largely reduced. This result implies that distributions and fluctuations for the local mixing ratio possibly induce inferior mechanical properties of small $f$ networks. In contrast, for $f = 8$ shown in the right column, the examined fracture characteristics are relatively insensitive to $\varphi_m$. This robustness of large $f$ networks is consistent with the widely attempted strategy for developing tough polymeric networks by employing multi-functional network nodes.

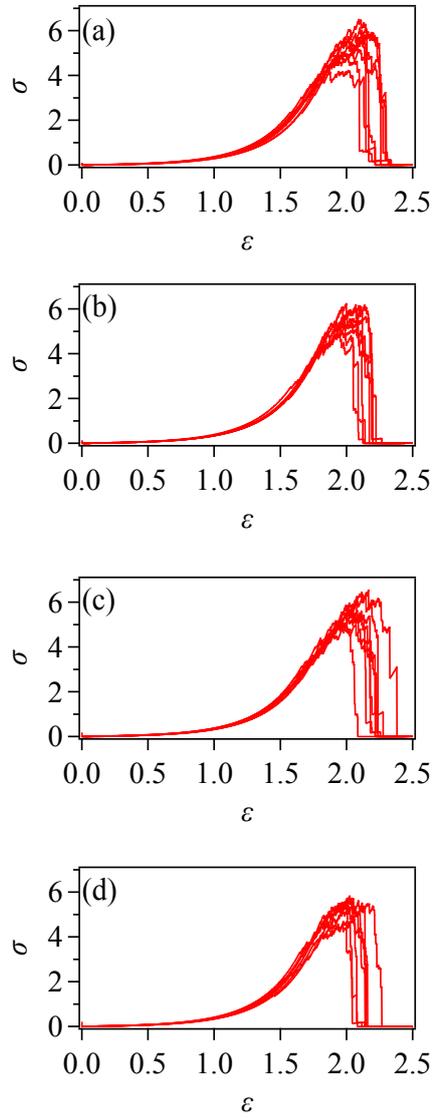

**Figure 1** Stress-strain curves for $(f, \varphi_c) = (4, 0.8)$ with $\varphi_m/(1 - \varphi_m) =$1, 1.1, 1.25, and 1.5 from top to bottom. Each curve exhibits the result of an independent simulation run.

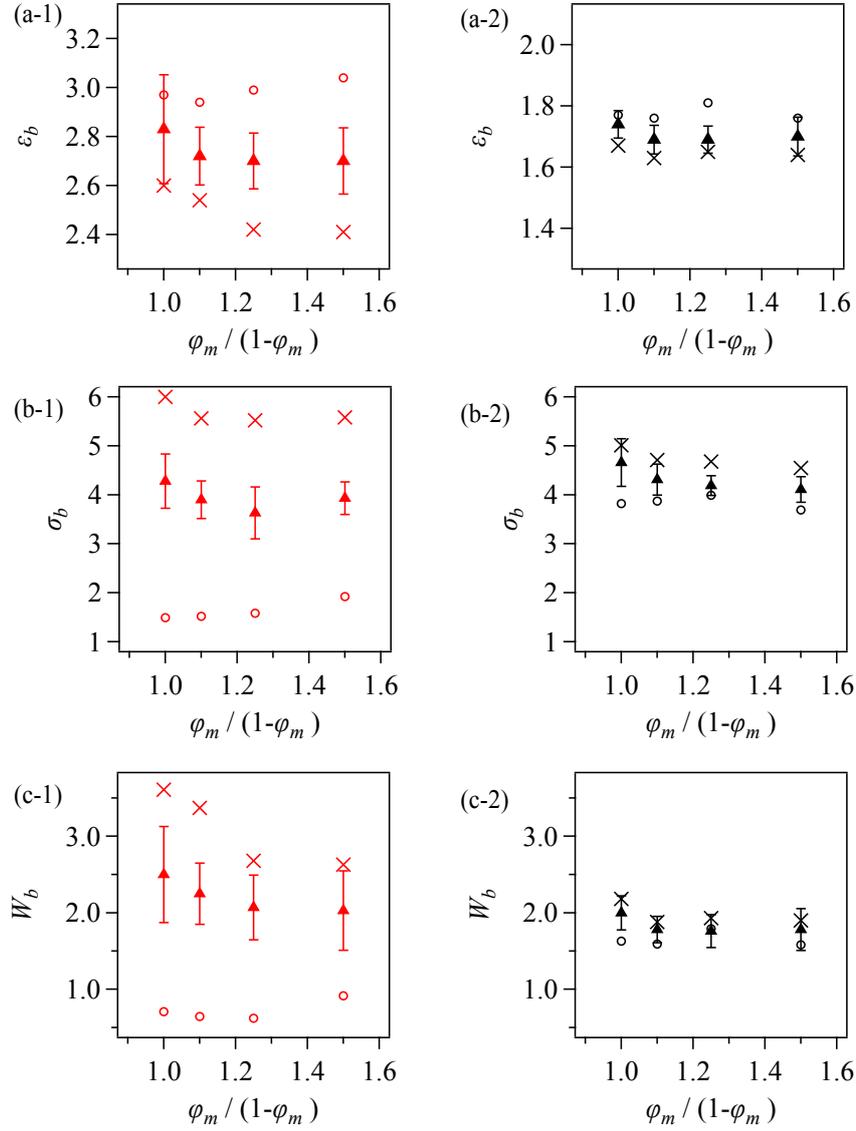

**Figure 2** Fracture characteristics $\varepsilon_b$, $\sigma_b$, and $W_b$ plotted against $\varphi_m/(1-\varphi_m)$ from top to bottom for $f=3$ (left) and 8 (right). Circle, triangle, and cross indicate the results at $\varphi_c=0.6, 0.7$ and 0.8, respectively. Error bars exhibit the standard deviations among eight independent simulation runs for $\varphi_c=0.7$.

The other important fracture characteristic is the number ratio of broken strands $\varphi_{bb}$, as discussed in the previous study [13]. Figure 3 exhibits this quantity for $f=3$ and 8 plotted against $\varphi_m/(1-\varphi_m)$. $\varphi_{bb}$ depends on the stoichiometry in a similar manner with $\sigma_b$ and $W_b$. Namely, for $f=3$, at high $\varphi_c$, $\varphi_{bb}$ decreases with increasing $\varphi_m/(1-\varphi_m)$, whereas for $f=8$, $\varphi_{bb}$ is insensitive to the stoichiometry.

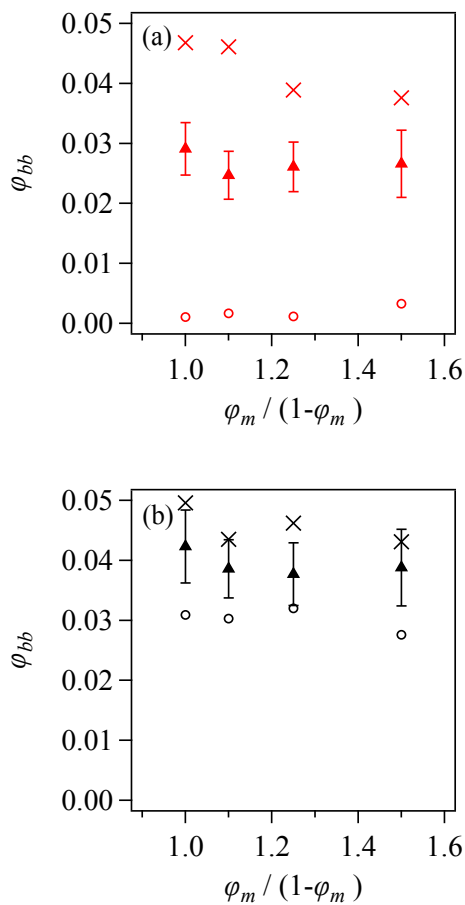

**Figure 3** $\varphi_{bb}$ for $f = 3$ (top) and 8 (bottom) at $\varphi_c = 0.6$ (circle), 0.7 (triangle), and 0.8 (cross) plotted against $\varphi_m/(1-\varphi_m)$. Error bars exhibit the standard deviations among eight independent simulation runs for $\varphi_c = 0.7$.

Figure 4 shows the cycle rank per prepolymer $\xi$ plotted against $\varphi_m$ for a few $(f, \varphi_c)$ cases. As made in the previous study,[12] $\xi$ was obtained from the number of effective network nodes and strands included in the percolated network, $\nu$ and $\mu$, as $\xi = \nu - \mu$. In this definition, $\xi$ does not depend on applying the Scanlan-Case criterion [30,31]. The obtained $\xi$ is close to the prediction of the mean-field theory [23–25] shown by the broken horizontal lines and depends on $\varphi_m$ only slightly. For the case of $f = 3$ (panel a), $\xi$ for $\varphi_c = 0.8$ (cross) weakly increases with increasing $\varphi_m/(1-\varphi_m)$. This increase of $\xi$ means an increase in the effective functionality of network nodes owing to the excess partner prepolymers. Although faint, the increase of $\xi$ is consistent with the decreasing $\varepsilon_b$, $\sigma_b$, and $W_b$ shown in Figure 2, since $\varepsilon_b$,

$\sigma_b$, and $W_b$ decreases with increasing $\xi$ [12].

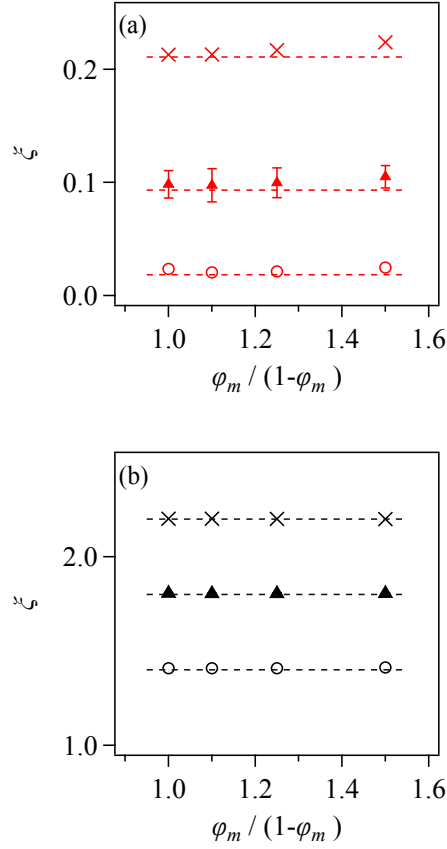

**Figure 4** $\xi$ for $f = 3$ (top) and 8 (bottom) at $\varphi_c = 0.6$ (circle), 0.7 (triangle), and 0.8 (cross) plotted against $\varphi_m/(1 - \varphi_m)$. Error bars indicate the standard deviations among eight independent simulation runs and are within symbols for $f = 8$. Broken lines are predictions of the mean-field theory.

Concerning the $\varphi_m$-dependence of $\xi$, let us briefly recapitulate the mean-field theory [23–25]. The probability for one of the prepolymer arms to be excluded from the percolated network $P_{out}$ is written as

$$P_{out} = \varphi_c P_{out}^{f-1} + (1 - \varphi_c) \qquad (1)$$

The first term on the right-hand side is the probability of a situation where the test arm is reacted, and the others connected to the same branch point are not included in the network. The second term is the unreacted probability for the test arm. One can obtain $P_{out}$ for a given set of $\varphi_c$ and $f$. For a $P_{out}$ thus obtained, the probability for a prepolymer (with functionality $f$) that has $h$ arms embedded in the network $P(h)$ is written as

$$P(h) = {}_fC_h(1 - P_{out})^h P_{out}^{f-h} \qquad (2)$$

With $P(h)$, the number of nodes and strands in the network, $\nu$ and $\mu$, can be written as below.

$$\nu = \sum_{i=3}^{f} iP(i) \qquad (3)$$

$$\mu = \sum_{i=3}^{f} P(i) \qquad (4)$$

Note that in eqs 3 and 4 the Scanlan-Case criterion [30,31] is applied. Finally, $\xi$ is calculated as $\xi = \nu - \mu$. In the calculations shown above, the reaction is assumed to take place independently on each arm, and it is not affected by the connectivity of neighboring prepolymers. Under such assumptions, the mixing ratio does not play any role, and indeed, it is not involved in the calculation of $\xi$ above. As shown in the previous study, for equimolar blends, the simulation results of $\xi$ are fully consistent with this calculation, demonstrating that correlations between reactions for different arms are negligible. In contrast, the $\varphi_m$-dependence of $\xi$ shown in Fig 4 (a) implies correlations between reactions for different arms, induced by the deviation from the equimolar condition, though not significant.

The stoichiometry dependence of the fracture characteristics appears consistent with the change of $\xi$. This argument is examined in Figure 5, which shows $\varepsilon_b$, $\sigma_b/\varphi_{bb}$, and $W_b/\varphi_{bb}$ as functions of $\xi$. As discussed above, the fracture characteristics depend on $\varphi_m$ when $f$ is small. Reflecting this result, the data points for $f \leq 4$ with different $\varphi_m$ values can be distinguished from each other, albeit within the error bars. Since the fracture characteristics for large $f$ values ($f \geq 5$) are insensitive to $\varphi_m$, the results of different $\varphi_m$ values mostly overlap with each other.

In Figure 5, the apparent relations obtained in the previous study [12] are also indicated for comparison by broken curves. The data obtained in this study lie on the master curves previously obtained. This result implies that the stoichiometry does not impact the network fracture. Concerning $\sigma_b/\varphi_{bb}$, the data would be better described by another relationship rather than the power-law expression, although no theory has been found.

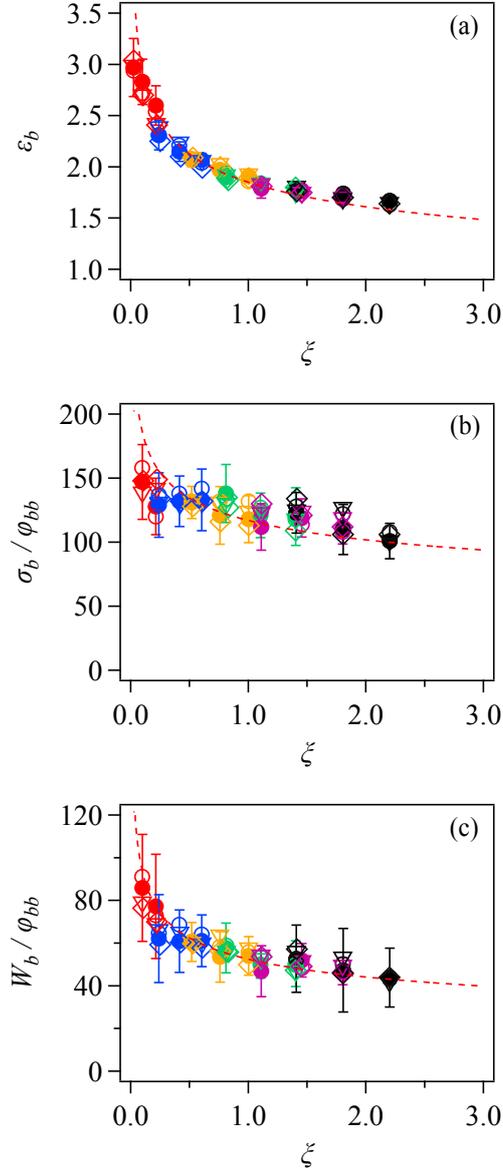

**Figure 5** $\varepsilon_b$, $\sigma_b/\varphi_{bb}$, and $W_b/\varphi_{bb}$ as functions of $\xi$ for $f=3$ (red), 4 (blue), 5 (orange), 6 (green), 7 (violet), and 8 (black) at $\varphi_c = 0.6$, 0.7, and 0.8, with $\varphi_m/(1-\varphi_m)=1$ (filled circle), 1.1 (unfilled circle), 1.25 (triangle), and 1.5 (square). Error bars show standard deviations for eight independent simulation runs for $\varphi_m/(1-\varphi_m)=1$. Broken curves exhibit the empirical relations $\varepsilon_b = 1.85\xi^{-0.2}$, $\sigma_b/\varphi_{bb} = 117\xi^{-0.2}$, and $W_b/\varphi_{bb} = 52.5\xi^{-0.25}$ reported previously. Note that error bars in the horizontal direction are within the symbols. Note also that the data for $(f,\varphi_c) = (3,0.6)$ are out of the plot ranges for panels (b) and (c).

It should be noted that the networks in the previous study do not include secondary loops, which are contained in the present simulations. Because the number of such loops is not large (less than

3% for $f=3$), they do not significantly impact the mechanical and structural properties examined here, although they are relatively unbroken.[20,26]

Concerning the above-reported universal relationships between fracture characteristics and $\xi$, prepolymers with only two of their arms reacted seem to play a role as strand extenders. Figure 6 shows the number-averaged strand molecular weight between network nodes with more than three reacted arms $M_n$ divided by the span molecular weight of the prepolymer $M_0$. At small $\xi$, there exist lots of strand extenders, and $M_n/M_0$ is close to 2, implying that almost all the effective nodes are connected by four prepolymer arms, not two. As $\xi$ increases, $M_n/M_0$ decreases to approach unity, reflecting that most prepolymers have more than 3 reacted arms. As reported previously, $M_n/M_0$ plotted against $\xi$ lie on a master curve for various $f$ values for the equimolar cases. Figure 6 demonstrates that the data for different mixing ratios also fall on the same curve. Since networks with longer strands are more stretchable, the observed universality for $\xi$-dependence of $M_n/M_0$ explains the results shown in Fig 5, at least partly.

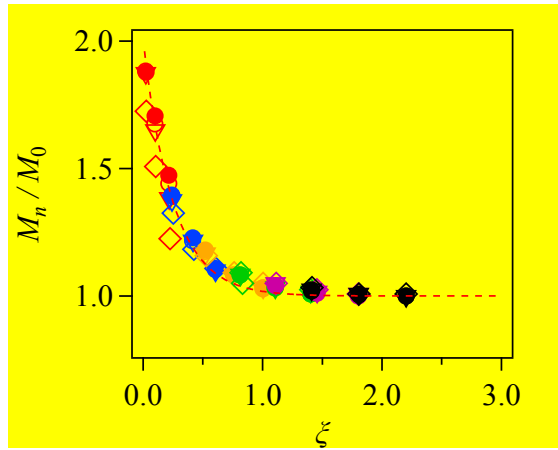

**Figure 6** Strand molecular weight $M_n$ between network nodes that have more than three reacted arms divided by the span molecular weight of the prepolymer $M_0$ as a function of $\xi$ for $f=3$ (red), 4 (blue), 5 (orange), 6 (green), 7 (violet), and 8 (black) at $\varphi_c = 0.6$, 0.7, and 0.8, with $\varphi_m/(1-\varphi_m)=1$ (filled circle), 1.1 (unfilled circle), 1.25 (triangle), and 1.5 (square). The broken curve exhibits the empirical relation, $M_n/M_0 = 1 + \exp(-4\xi)$.

## CONCLUSIONS

The phantom chain simulations were performed on the fracture of polymer networks made from binary mixtures of star prepolymers with various mixing ratios. In the obtained stress-strain relations, the fracture characteristics, including strain and stress at break and work for fracture, are weakly dependent on stoichiometry when the node functionality is small, and the conversion ratio is high. At the same time, they are insensitive to the stoichiometry otherwise, reflecting that the end-linking reactions are not strongly correlated. The ratio of broken bonds and the cycle rank of the networks exhibit similar behavior. Consequently, the stoichiometry ratio does not disturb the relations between the cycle rank and the fracture characteristics reported previously. These results suggest that the stoichiometry of prepolymer mixtures does not impact the fracture properties of star polymer networks.

Further studies are ongoing on the effect of molecular weight distribution of prepolymers[19,26], mixing of polymers with different functionalities[10,32], entanglements[27,33], etc.; the results will be reported elsewhere.

## ACKNOWLEDGEMENTS

This study is partly supported by JST-CREST (JPMJCR1992) and JSPS KAKENHI (22H01189).